\documentclass{aa}

\usepackage[utf8]{inputenc}
\usepackage{lmodern}
\usepackage[T1]{fontenc}
\usepackage{url}
\usepackage{soul}
\usepackage[pdftex,dvipsnames,hyperref]{xcolor}
\usepackage{graphicx}
\usepackage{txfonts}
\usepackage{natbib}

\makeatletter
\renewenvironment{table}
{\@float{table} \small}
{\end@float}
\makeatother

\begin{document} 

   \title{Sparse aperture masking at the VLT \thanks{Based on observations collected at the European Southern Observatory (ESO) during runs 087.C-0450(A), 087.C-0450(B) 087.C-0750(A), 088.C-0358(A).}}
   \subtitle{II. Detection limits for the eight debris disks stars $\beta$ Pic, AU Mic, 49 Cet, $\eta$ Tel, Fomalhaut, g Lup, HD181327 and HR8799 }

   \author{L. Gauchet
          \inst{1}\thanks{Email: \url{lucien.gauchet@obspm.fr}}
          \and
          S. Lacour\inst{1}
          \and
          A.-M. Lagrange\inst{2,3}
          \and
          D. Ehrenreich \inst{4}
          \and
           M. Bonnefoy\inst{2,3}
          \and
          J. H. Girard\inst{5}
          \and
          A. Boccaletti\inst{1}
          }

    \institute{LESIA, Observatoire de Paris, PSL Research University, CNRS, Sorbonne Universités, Univ. Paris Diderot, UPMC Univ. Paris 06, Sorbonne Paris Cité, 5 place Jules Janssen, F-92195 Meudon, France
    \and
    Univ. Grenoble Alpes, IPAG, F-38000 Grenoble, France 
    \and 
    CNRS, IPAG, F-38000 Grenoble, France
    \and
    Observatoire de l’Université de Genève, 51 chemin des Maillettes, 1290 Sauverny, Switzerland
    \and
    European Southern Observatory, Casilla 19001, Santiago 19, Chile
            }

   \date{Received ... ; accepted ...}

    \abstract
    {The formation of planetary systems is a common, yet complex mechanism. Numerous stars have been identified to possess a debris disk, a proto-planetary disk or a planetary system. The understanding of such formation process requires the study of debris disks. These targets are substantial and particularly suitable for optical and infrared observations. Sparse Aperture masking (SAM) is a high angular resolution technique strongly contributing to probe the region from 30 to 200 mas around the stars. This area is usually unreachable with classical imaging, and the technique also remains highly competitive compared to vortex coronagraph.}
    {We aim to study debris disks with aperture masking to probe the close environment of the stars. Our goal is either to find low mass companions, or to set detection limits.}
    {We observed eight stars presenting debris disks ( $\beta$ Pictoris, AU Microscopii, 49 Ceti, $\eta$ Telescopii, Fomalhaut, g Lupi, HD181327 and HR8799) with SAM technique on the NaCo instrument at the VLT.}
    {No close companions were detected using closure phase information under $0.5''$ of separation from the parent stars. We obtained magnitude detection limits that we converted to Jupiter masses detection limits using theoretical isochrones from evolutionary models.}
    {We derived upper mass limits on the presence of companions in the area of few times the diffraction limit of the telescope around each target star.\thanks{All magnitude detection limits maps are only available in electronic form at the CDS via anonymous ftp to cdsarc.u-strasbg.fr (130.79.128.5) or via \url{http://cdsweb.u-strasbg.fr/cgi-bin/qcat?J/A+A/}} }

    \titlerunning{Detection limits in debris disks around young stars with NaCo/SAM}
    \maketitle
%
\section{Introduction}

\begin{table*}
    \centering
    \caption[]{Log of SAM observations with VLT/NACO.}
    \label{ObsLog}
    \begin{tabular}{lllllllll}
        \hline \hline \noalign{\smallskip}
        Star & UT date & Mask & DIT  & NDIT & NEXP & Seeing$^{(1)}$ & $\tau_0^{(2)}$ & Calibrator \\
        & y/m/d & & (s) & & & & (ms) & \\
        \noalign{\smallskip} \hline \noalign{\smallskip}
        g Lup & 2011/06/08 & 7 holes & 0.08 & 500 & 40 & 0.76 & 3.2 & HD139960 \\
        49 Cet & 2011/08/04 & 7 holes & 0.08 & 500 & 40 & 0.67 & 1.5 & HD10100 \\
        Fomalhaut & 2011/08/04 & BB 9 holes & 0.02 & 2000 & 32 & 0.68 & 2.1 & del PsA \\
        HD181327 & 2011/08/04 & 7 holes & 0.106 & 472 & 40 & 0.80 & 2.3 & HD180987 \\
        HR8799 & 2011/08/04 & 7 holes & 0.051 & 785 & 40 & 0.77 & 1.6 & HD218234 \\
        AU Mic & 2011/08/31 & 7 holes & 0.04 & 600 & 104 & 0.82 & 3.1 & HD197339 \\
        $\eta$ Tel & 2011/09/01 & 7 holes & 0.01 & 500 & 56 & 0.96 & 2.1 & HD181517 \\
        $\beta$ Pic & 2011/10/12 & 7 holes & 0.04 & 800 & 64 & 1.8 & 0.6 & HR2049 \\
        \noalign{\smallskip} \hline
        \multicolumn{9}{l}{ \tiny $^{(1)}$ provided using keyword \texttt{TEL IA FWHM} $^{(2)}$  provided using keyword \texttt{TEL AMBI TAU0}}\\
    \end{tabular}
\end{table*}

Disk evolution is a key question in the comprehension of planetary system formation as they are intrinsically linked together. Thus direct detection and imaging of planets in such disks are mandatory to understand how planetary systems form and evolve. A debris disk is composed of dust grains and small bodies orbiting around the star, it is usually gas-poor, and grain sizes are typically one to hundred of micrometers. The debris disks depletion process is recognized to be short ($\sim$10Myr or less) but many disks have been found to last longer. The presence of large planetary bodies can induce sufficient gravitational perturbations to provoke collisions of planetesimals, the result is smaller fragments that replenish the circumstellar environment.
Thus debris disks may be good indicators of planetary systems or ongoing planet formation.

The disks can be observed thanks to scattered light and thermal emission of the dust grains. Identification of infrared (IR) excess in the spectral energy distribution led to the first extra-solar debris disk detection around Vega \citep{aumann1984}. 

$\beta$ Pictoris is a remarkable system because of its young age estimated between 8 and 20 Myr \citep{zuckerman2001}. The existence of a planet \citep{lagrange2010} orbiting around $\beta$ pictoris makes it a benchmark for testing planetary formation models. Several scenarios were formalized, such as accretion, gravitational instabilities, or a combination of both \citep{bonnefoy2013}. It is decisive to understand which processes are involved in such formation and estimate to what extent they contribute to it.
Other young planets exist (\textit{e.g.} HR8799). It is thus utterly important to find other planets or debris disks stars to multiply reference points.

This paper aims to study eight objects around which debris disks have already been discovered. It is necessary to investigate the close environment of the star that is usually inaccessible to imaging techniques like coronagraphy and PSF subtraction algorithms (ADI, LOCI, PCA, etc.). Sparse Aperture Masking (SAM) has a small inner working angle, and has a unique ability to probe the region between $\frac{\lambda}{2D}$ and $\frac{2\lambda}{D}$, with high contrast capabilities (with a dynamic of a few thousands). It includes the area from 30 to 200 mas, which is equivalent to a distance of 1.5 to 10 AU for a star located at 50 pc, as for $\beta$ Pic, $\eta$ Tel, HR8799 and AU Mic. These separations correspond to regions in primordial disks where planets can efficiently form by core accretion \cite{kennedy2008}.

In Section \ref{sectionObs} we describe aperture masking observations performed on NaCo at the Very Large Telescope (VLT) and the data reduction technique, in Section \ref{sectionResults} we report our observational results in terms of mass limits of companions for each systems. In Section \ref{sectionConclusion} we present our conclusions.


\section{Observations and data reduction}
\label{sectionObs}

\subsection{Observations}
\label{subsectionObs}

The eight systems imaged, among which $\beta$ Pictoris and HR8799, are listed in Table \ref{ObsLog}. They were observed at the VLT from June to October 2011 using NAOS-CONICA (NaCo) providing adaptive optics assisted imaging.

We used the Sparse Aperture Mode (SAM) to achieve the highest angular resolution at the diffraction limit. This technique described in \cite{tuthill2000} uses a mask with holes in non-redundant configuration placed in a pupil plane of the instrument. The goal of this mask is to transform the main pupil of the telescope into an interferometric array. Each baseline made of any pair of sub-apertures will create a fringe pattern that is unique in terms of direction and spatial frequency. This allows to overcome the  non-coherent addition of the wavefront in the focal plane.

The observations were carried out in the $L'$-band ($\lambda_c=3.80\, \mu m$, $\Delta\lambda=0.62\, \mu m$) using two of the four masks available on NaCo, the "Broad Band (BB) 9 holes" and the "7 holes" masks. Observational details such as mask used, integration time (DIT), size of datacube, and number of exposures for each science target are summarized in Table \ref{ObsLog}.

\begin{figure}
    \centering
    \resizebox{\hsize}{!}{\includegraphics[width=0.24\textwidth]{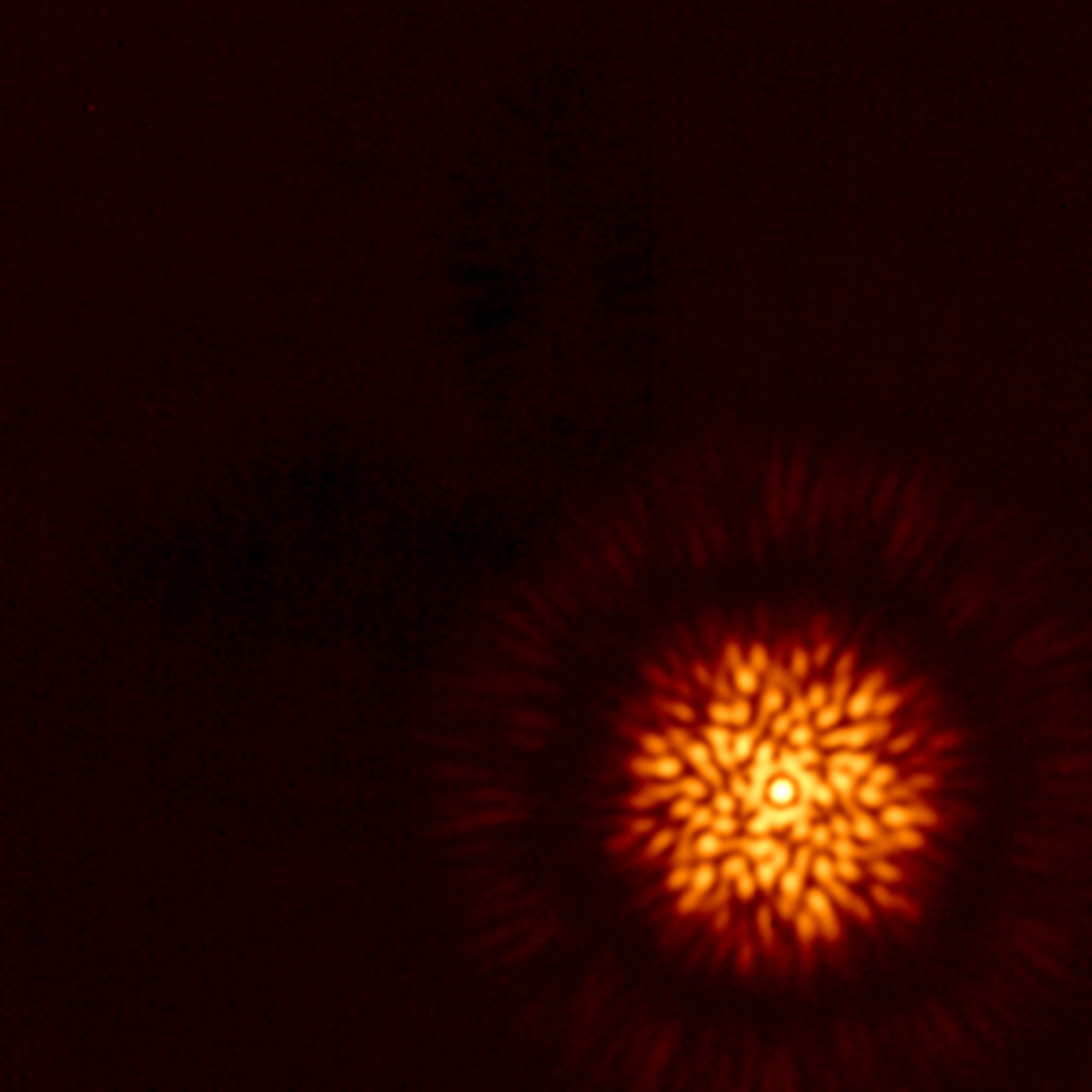} \includegraphics[width=0.24\textwidth]{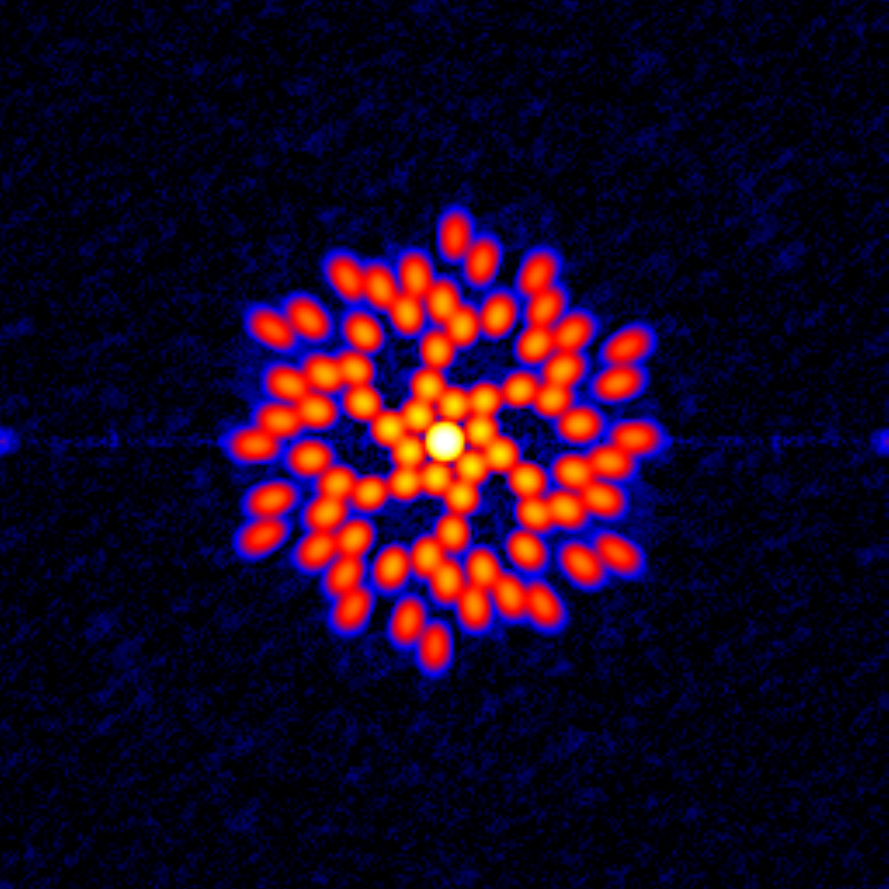} }
    \caption{\textbf{Left:} Image of Fomalhaut on the $256\times256$-pixel CONICA detector. The detector field-of-view is $7^{''}$; i.e., each pixel is 27 mas in size on sky. The $\sim2.6$ arcsec point spread function is due to diffraction of the 78 cm holes in the 9-holes mask used during this observation. The many speckles apparent inside this large Airy pattern are due to the superposition of fringes caused by the 36 spatial frequencies allowed through the mask. \textbf{Right:} Power spectrum \textit{i.e.} squared amplitude of Fourier's components from the Fomalhaut PSF in the spatial frequency domain.}
    \label{fig:FomalhautPSF}
\end{figure}

The operational mode for observation was to image the star with the masked pupil in each of the four quadrants of the L27 camera of NaCo (the four $128\times128$ pixels quadrants from the $256\times256$ pixels detector). An example of PSF and respective power spectrum obtained with Fomalhaut are displayed on Figure~\ref{fig:FomalhautPSF}. Each target is alternately observed with a calibrator star. This provides a PSF reference in order to calibrate the science data. This process benefits from the SAM "Star Hopping" strategy allowing a fast switch between science star and calibrator. As long as the two targets have comparable brightness on the wavefront sensor it is possible to jump from one to the other without requiring a full optimization of the AO loop and its associated time penalty. Thus we get very similar atmospheric perturbation over both science and respective calibrator's data. This significantly improves accuracy on calibration and results in higher performance over detection limits.

\subsection{SAM pipeline}

The data was reduced using the Sparse Aperture Mode Pipeline (SAMP) written in the Yorick interpreted language. Description and operating details of the pipeline are available in \cite{lacour2011}. Here we summarize the main method of reduction. In the first stage, data is corrected from systematic alterations: images are flat-fielded, bad-pixels removed, sky-subtracted. This step is particularly important as the sky luminosity is dominant in $L'$-band. Quadrant images from the 8-point offset pattern are aligned, centered and stacked. Direct fringes fitting is then performed to get complex visibilities from the fringes \citep[see][]{greenbaum2015}. Amplitude and phase of the fringes are strongly affected by the atmospheric turbulence. To overcome perturbations from the atmosphere, we use the closure phase quantity (\textit{e.g.} \citealt{baldwin1986, haniff1987}) which takes into account the symmetry of the object. It consists in the linear combination of phases over a triangle of baselines. This quantity has the interesting property of being independent from the atmospheric piston. More precisely, complex visibilities are multiplied together to form the triple product called bispectrum. Closure phases are obtained from the argument of the bispectrum \citep{weigelt1977,lohmann1983}. All closure phases are calibrated by subtraction of the closure phases from the calibrator stars mentioned in Section \ref{subsectionObs}.

The observed closure phases are then fitted with a binary model with three parameters: separation $\delta$, position angle $\alpha$ and luminosity ratio $\rho$. A cube of $\chi^2$ values from each triplet $(\alpha, \delta, \rho)$ is built. Since no clear detection is present in our dataset, the $\chi^2(\alpha, \delta, \rho)$ cube is scaled so that the reduced $\chi^2$ is equal to 1 for $\rho = 0$ (no companion). Detection limits are then obtained for each $\alpha$ and $\delta$ as $\tilde \rho$ such as $\chi^2(\alpha, \delta, \tilde \rho) < 25 + \chi^2(0, 0, 0)$. In other words, the detection limits are calculated as the minimum value for which the $\chi^2$ is below 5~$\sigma$ of a non detection.

\begin{table*}
    \caption[]{List of stars used in this work, provided with specifications: distance from the observer, estimated age and spectral type of the parent star.}
    \label{listObjects}
    \centering
    \begin{tabular}{llllllc} 
        \hline \hline \noalign{\smallskip}
        Object & Distance & Age  & Spectral Type & K$^{(1)}$ & $L'$$^{(2)}$ & References\\
            & (pc) & (Myr) & & (mag) & (mag) & \\
        \noalign{\smallskip} \hline \noalign{\smallskip}
        49 Cet & 59 & 40 & A1V & 5.46 & 5.45 & 1,2\\
        AU Mic & 9.9 & 21\raisebox{.35ex}{$\scriptstyle\pm 4$} & M1Ve & 4.53 & 4.32 & 3,4,5\\ 
        $\beta$ Pic & 19.3 & 21\raisebox{.35ex}{$\scriptstyle\pm 4$} & A6V & 3.48 & 3.46 & 5,6 \\ 
        $\eta$ Tel & 47.7 & 21\raisebox{.35ex}{$\scriptstyle\pm 4$} & A0V & 5.01 & 5.01 & 1,5,7 \\ 
        Fomalhaut & 7.7 & 440\raisebox{.35ex}{$\scriptstyle\pm 40$} & A3V/A4V & 1.05 & 1.04 & 6,8 \\
        g Lup & 17.5 & $300^{+700}_{-200}$ & F5V & 3.80 & 3.76 & 1,9,10\\
        HD181327 & 51.8 & 21\raisebox{.35ex}{$\scriptstyle\pm 4$} & F5V/F6V & 5.91 & 5.87 & 1,5,11,12\\ 
        HR8799 & 39.4 & $30^{+20}_{-10}$ & F0V & 5.24 & 5.22 & 2,13 \\
        \noalign{\smallskip} \hline
        \multicolumn{7}{l}{ \tiny $^{(1)}$ provided by SIMBAD database}\\
        \multicolumn{7}{l}{ \tiny $^{(2)}$ derived with \cite{tokunaga2000} color table}
    \end{tabular}
    \tablebib{(1) \citet{cutri2003};
        (2) \cite{vanleeuwen2007}; 
        (3) \citet{stauffer2010};
        (4) \citet{torres2006};
        (5) \citet{binks2014};
        (6) \citet{ducati2002};
        (7) \citet{wyatt2007};
        (8) \citet{mamajek2012};
        (9) \citet{gray2006};
        (10) \citet{kalas2006};
        (11) \citet{schneider2006};
        (12) \citet{holmberg2009};
        (13) \citet{gray2014}.
        }
\end{table*}

\subsection{Detection limits: from magnitude to mass}
\begin{figure}
    \centering
    \resizebox{\hsize}{!}{\includegraphics[width=0.5\textwidth]{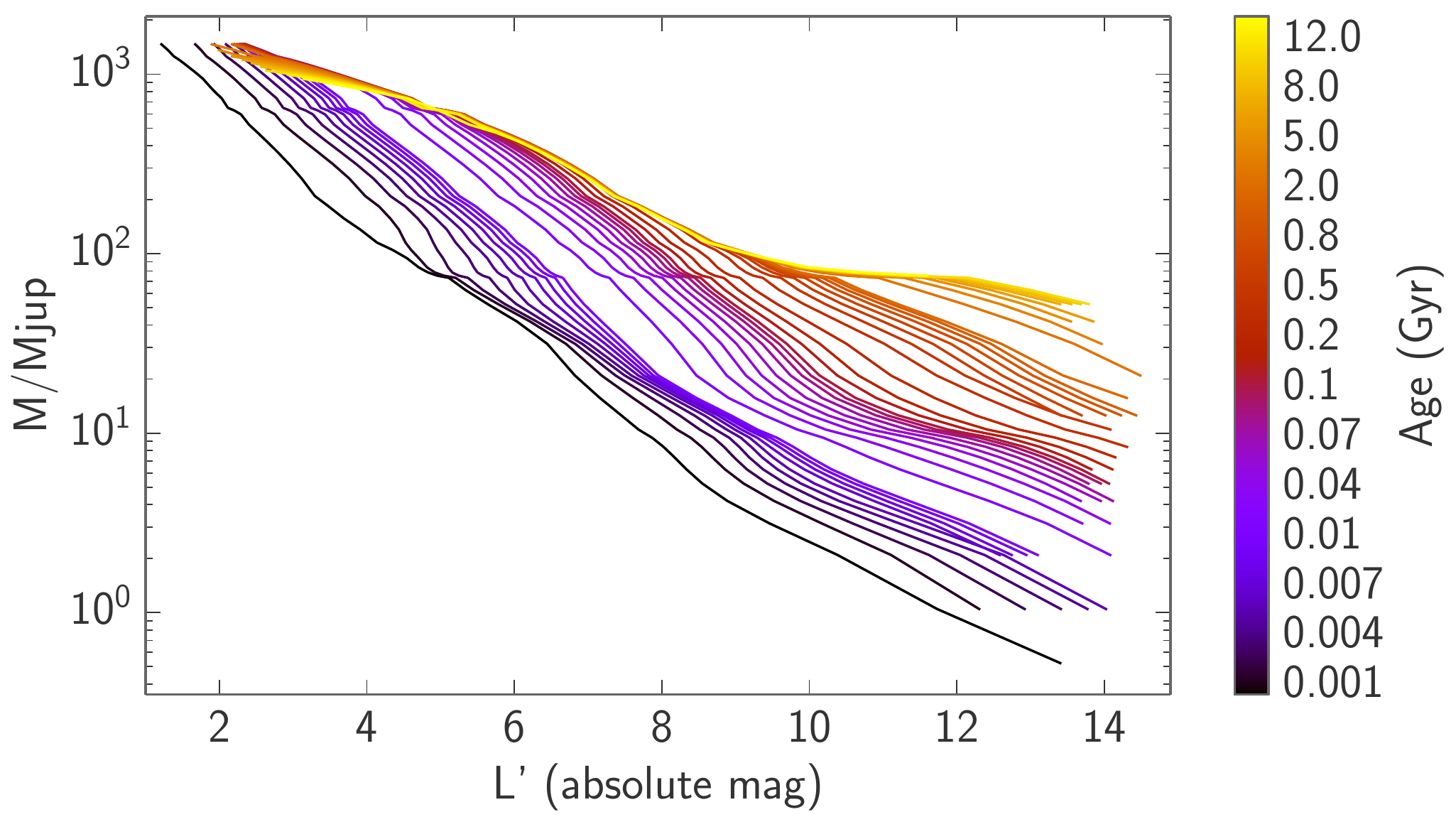}} 
    \caption{Theoretical isochrones showing $L'$ emission magnitude with mass of the object. }
    \label{fig:isochrones}
\end{figure}

\begin{figure}
    \centering
    \resizebox{\hsize}{!}{\includegraphics[width=0.5\textwidth]{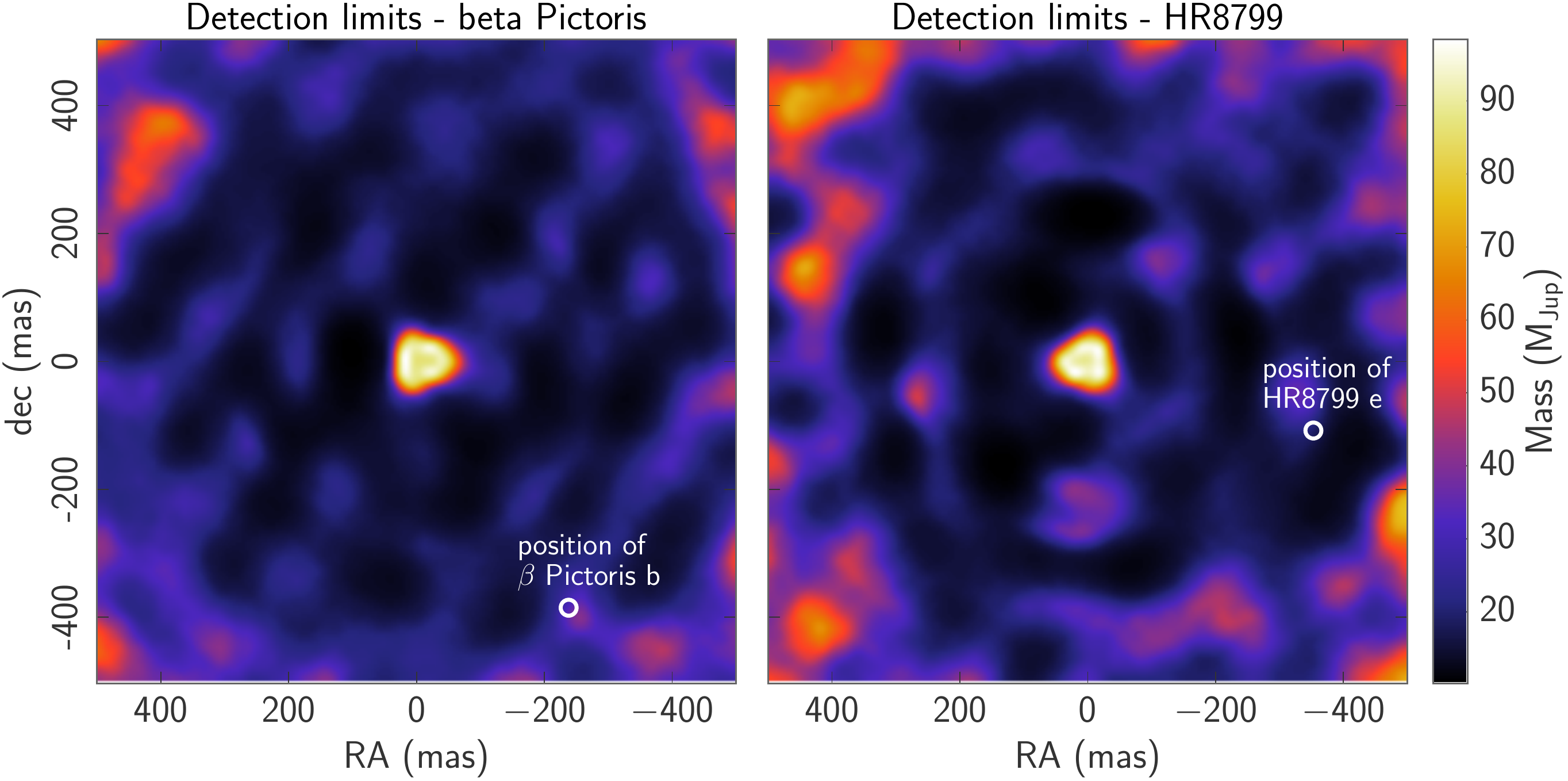}}
    \caption{Sample of detection limits maps obtained for $\beta$ Pictoris (Left) and HR8799 (Right)}
    \label{fig:LODMaps}
\end{figure}

The examination of closure phases did not allow to rule on the presence of a faint companion in the data sets. However, study of the $\chi^2$ maps allows for the extraction of detection limits. 

Detection limit maps from the \emph{SAM pipeline} are expressed in contrast ratio. Hence, the $L'$ band detection limits depend on the brightness of the parent star. We used the spectral type of the stars (given in Table \ref{listObjects}), combined with \cite{tokunaga2000} tables of intrinsic colors for main sequence, giant and super giant stars, to derive the $L'$ absolute magnitude of the stars. The magnitudes are listed in the Table \ref{listObjects}. At this point, we obtain detection limits in term of absolute $L'$ magnitude\footnote{Detection limits maps are available in electronic form at the CDS}.

To estimate the mass as a function of the luminosity, we used the theoretical isochrones established by \cite{baraffe1997, baraffe1998, baraffe2003}. More specifically, we used the BT-Settl Model \citep{allard2012}. Those isochrones were interpolated into synthetic color tables and converted into the VLT NaCo filters systems\footnote{\url{https://phoenix.ens-lyon.fr/Grids/BT-Settl/}}.
Other evolutionary models do exist and are based on similar or alternative planet formation scenarios, such as the 'warm-start' model proposed by \cite{spiegel2012}. The mass/luminosity isochrones derived from other models can be perfectly applied to the data we made available online.
Figure~\ref{fig:isochrones} displays the theoretical isochrones for $L'$ magnitude versus the mass of the object. Given the commonly accepted age of the parent star/debris disk and the absolute magnitude of the detection limit, we derive the associated mass. Examples of two dimensional mass detection limit maps for $\beta$ Pictoris and HR8799 are displayed on Figure~\ref{fig:LODMaps}.

\section{Results}
\label{sectionResults}

\begin{figure*}
    \centering        
    \resizebox{0.86\hsize}{!}{\includegraphics[width=0.8\textwidth]{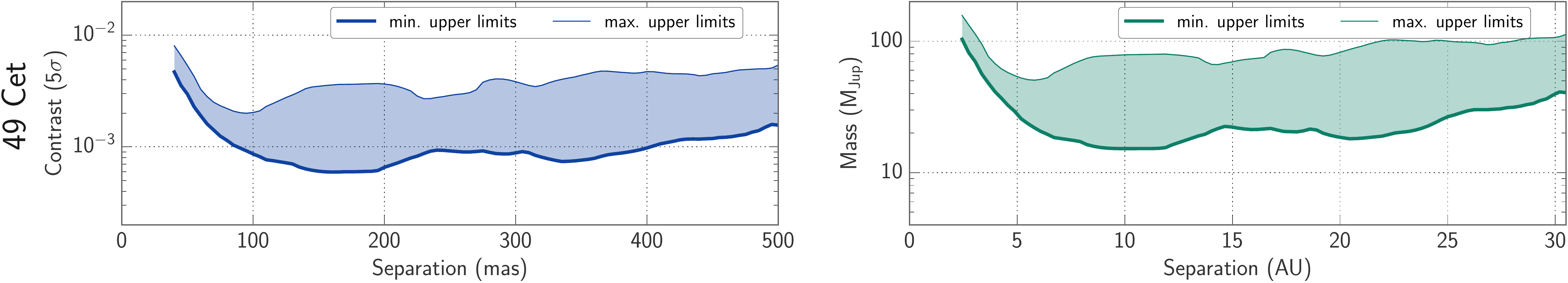}}
    \resizebox{0.86\hsize}{!}{\includegraphics[width=0.8\textwidth]{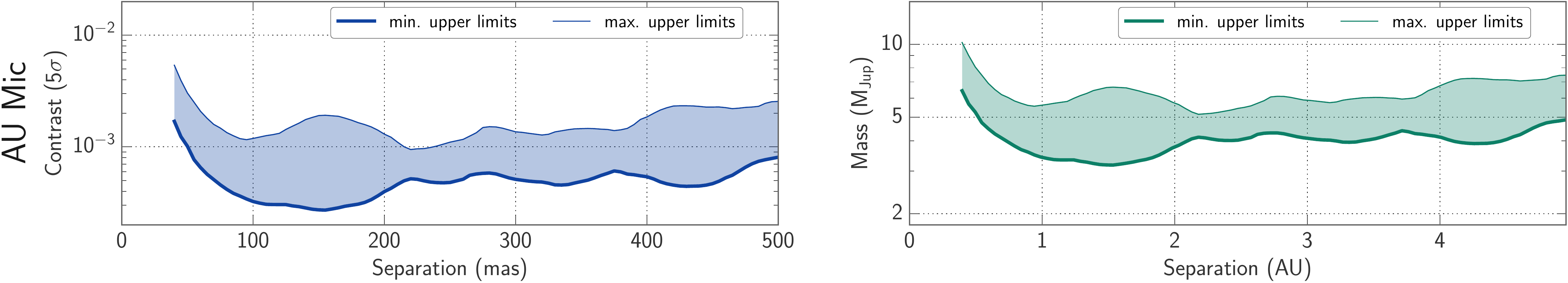}}
    \resizebox{0.86\hsize}{!}{\includegraphics[width=0.8\textwidth]{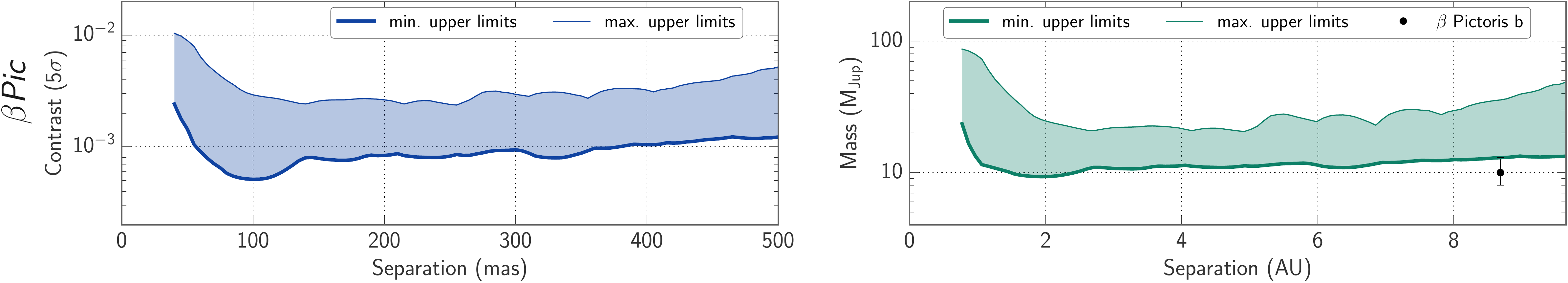}}
    \resizebox{0.86\hsize}{!}{\includegraphics[width=0.8\textwidth]{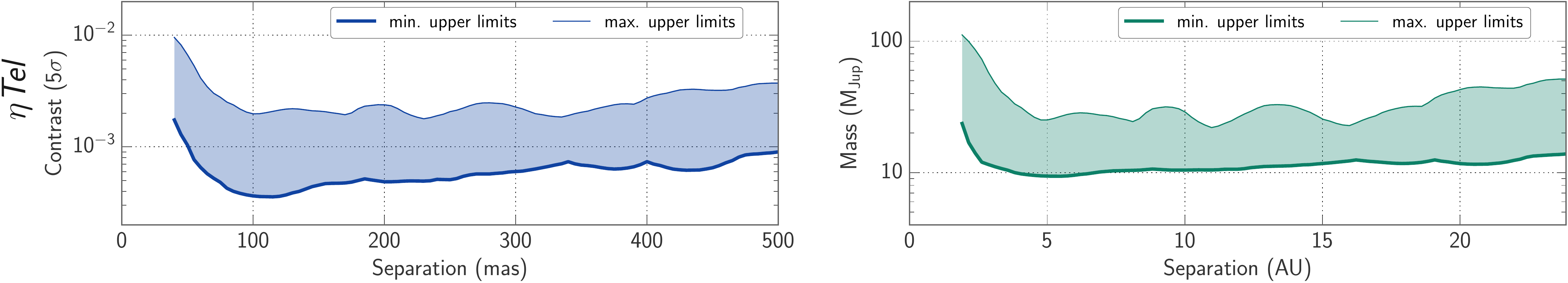}}
    \resizebox{0.86\hsize}{!}{\includegraphics[width=0.8\textwidth]{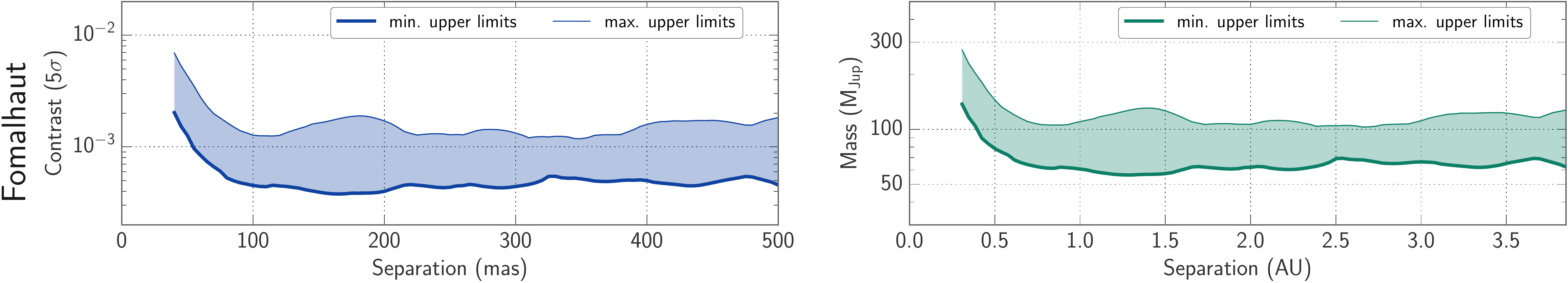}}
    \resizebox{0.86\hsize}{!}{\includegraphics[width=0.8\textwidth]{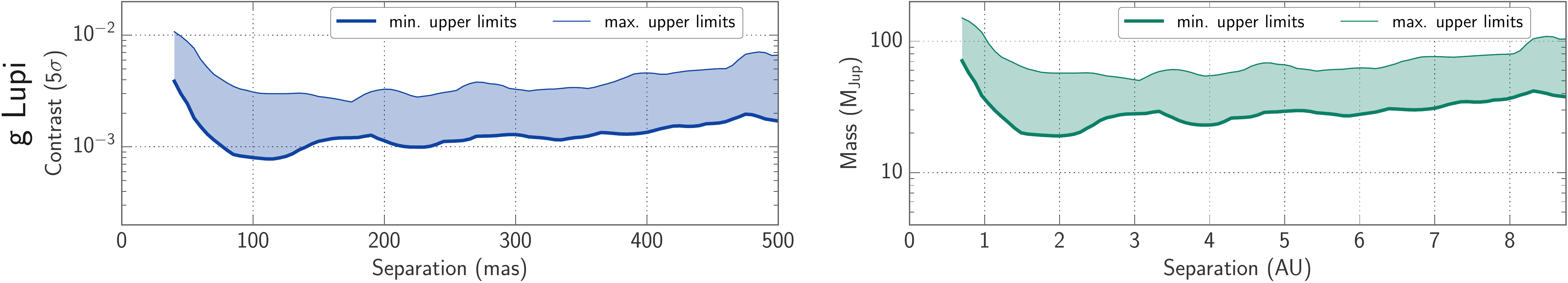}}
    \resizebox{0.86\hsize}{!}{\includegraphics[width=0.8\textwidth]{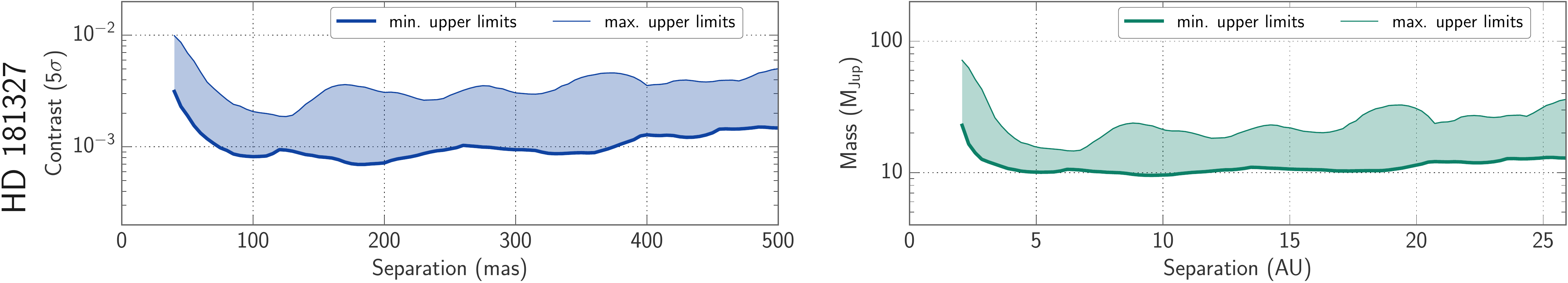}}
    \resizebox{0.86\hsize}{!}{\includegraphics[width=0.8\textwidth]{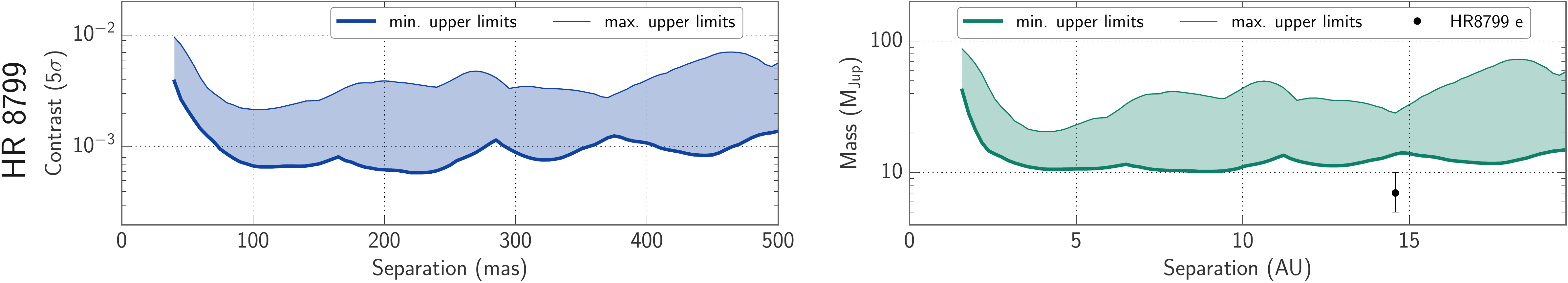}}
    
    \caption{\textbf{Radial distribution of the detection limits}:
    On the left, blue graphs display contrast as a function of the angular distance in milliarcseconds, this expresses the performance of the detection. On the right, green graphs display estimated mass detection limits (according to the BT-Settl model) as a function of separation in AU. In order to keep track of the azimuthal variation, the lower thick line represents the minimum value found at a given radial distance $r$, while upper thin line is the maximum value. Black dots represent companion within the known planetary systems.
    }
    \label{fig:LODRadial}
\end{figure*}

From the two-dimension maps, we determine a radial distribution of the detection limits. It is obtained by computing the minimum and the maximum value in 5 mas thick rings. The computation was done for radii between 30 to 500 mas. This process allows to keep the information on variations over azimuthal position at a given radius. The radial detection limits for each target are displayed in Figure~\ref{fig:LODRadial}.  Table~\ref{SAMContrast} summarizes the highest contrast achieved in the 30-200\ mas separation range -- where SAM is the most efficient. The $\Delta mag$ value is obtained by taking the median of the highest contrast values within the afore mentioned separation range. This median makes a good estimation of the bottom plateau value.
Finally, Table \ref{SAMUpperMasLimits} displays the upper mass limit for each object, \textit{i.e.} the lowest mass we can safely exclude over the whole 30 to 500 mas range.

\subsubsection*{\object{49 Ceti}}
49 Ceti is a main sequence A1V star identified to be a member of the 40 Myr old Argus association \citep{zuckerman2012}. It is located at a distance of $59\pm1$pc \citep{vanleeuwen2007}. This disk is unusual because of its high density of gas, such as CO, compared to dust. It indicates a non primordial gas presence likely to come from comet-like planetesimals collisions \citep{roberge2013}.

Detection limits at $5\sigma$ on 49 Ceti give at least a contrast of $5\cdot 10^{-3}$ beyond 50\ mas, up to $2\cdot 10^{-3}$ at 95\ mas separation ($\Delta mag$ of 5.8 and 8 respectively).

\subsubsection*{\object{AU Microscopii}}
AU Mic is a young $0.3 M_{\sun}$ M dwarf \citep{plavchan2009}. Its age is estimated to be 8-20 Myr, and its Hipparcos
distance is 9.9pc \citep{perryman1997}. It belongs to the $\beta$ Pictoris moving group. It features a debris disk seen nearly edge-on \citep{kalas2004}.

The youth and proximity of AU Mic are really favorable conditions to look for a low mass detection domain. Recently, fast moving ripples in the disk has been observed. The ripples can be interpreted has signposts of significant planetary formation activity \citep{boccaletti2015}.

Detection limits at $5\sigma$ on AU Microscopii give at least a contrast of $2.5\cdot 10^{-3}$ beyond 50\ mas, up to $9.5\cdot 10^{-4}$ around 220\ mas ($\Delta mag$ of 6.5 and 7.6 respectively).

\subsubsection*{\object{Beta Pictoris}}
$\beta$ Pictoris is an A6V star located at 19.3\ pc \citep{vanleeuwen2007}. It belongs to the eponymous moving group, that was assumed to be $12_{-4}^{+8}$ Myr old \citep{zuckerman2001}. Recently, a more precise -- and possibly more accurate -- measurement of the lithium depletion boundary gave an age for the moving group of $21\pm4$ Myr \citep{binks2014}. 
The $\beta$ Pictoris system was the first optically resolved disk \mbox{\citep{smith1984}} and possesses a directly imaged exoplanet with one of the closest estimated semi-major axis to date: $\beta$ Pictoris b  \citep{lagrange2010,chauvin2012,bonnefoy2014}.

Considering the position of $\beta$ Pictoris b, it is within the field of view of NaCo/SAM. Despite the absence of detection through closure phases, we can constrain an upper $L'$ band contrast ratio -- and mass limit -- at that position. Thanks to the orbital characterization by \cite{chauvin2012} and concomitant observations by \cite{nielsen2014}, we estimated $\beta$ Pictoris b  to be situated at position $\Delta\alpha=-0.238''$ and $\Delta\delta=-0.385''$ during the observations reported here. The position is displayed in figure~\ref{fig:LODMaps}. The corresponding mass in terms of limits of detection (LOD) is $m_{LOD} = 30\;M_{Jup}$ ($\Delta mag=6.2$). This value is consistent with the non-detection of $\beta$ Pic b which has an estimated mass of $10^{+3}_{-2}\;M_{Jup}$ \citep{bonnefoy2014}.

Detection limits at $5\sigma$ on $\beta$ Pictoris b give a rather constant contrast between $5\cdot 10^{-3}$ and  $2.5\cdot 10^{-3}$ ($\Delta mag$ of 5.8 and 6.5 respectively) beyond 70mas.

\subsubsection*{\object{Eta Telescopii}}

Eta Telescopii (HR7329A) is an A0V star. It is located at 47.7 pc according to \citet{wyatt2007} and is also a member of the $\beta$ Pictoris Moving Group. It possesses a companion HR7329B at a 4 arsecond separation, out of the field of view of NaCO/SAM instrument. $\eta$ Tel has also a more distant stellar companion HD181327 at a 7 arcmin separation.

Detection limits at $5\sigma$ on $\eta$ Telescopii give at least a contrast varying between $4\cdot 10^{-3}$ and $1.8\cdot 10^{-3}$ ($\Delta mag$ of 6 and 6.9 respectively) beyond 50mas.

\subsubsection*{\object{Fomalhaut}}
Fomalhaut is a A3V/A4V star located at only 7.7\ pc from the Sun. 
Fomalhaut's dust disk was initially discovered because of its infrared excess in IRAS observations by \cite{aumann1985}. Fomalhaut's dust belt was optically resolved by \cite{kalas2005} using the Advanced Camera for Surveys (ACS) on the Hubble Space Telescope (HST). The first optical image of an extra-solar planet around this system was obtained with the HST later on by \cite{kalas2008}.
Contrary to AU Mic or $\beta$ Pic, Fomalhaut's disk is not seen edge on, so it provides a better insight of the innermost region.

The very good seeing during the observation allowed to reach a 5$\sigma$ contrast ratio limit of $4.4 \times 10^{-4}$ within 50 to 200\,mas. However the relatively old age of Fomalhaut \citep[440 Myr,][]{mamajek2012} explains the high masses limit we obtain, since evolutionary models predict a decrease in luminosity with age.

A compilation of measurements from literature was done by \cite{kenworthy2013}. It gathered coronographic data from Spitzer ($4.5\ \mu m$) and VLT/NaCo ($4.05\ \mu m$) as well as VLTI/PIONIER interferometric data. Those measurements cover a large range of angular separation ($1$ to $2000\ mas$), nevertheless a gap in coverage remained from 100 to 200 mas. \citeauthor{kenworthy2013} pointed out in their Discussion section that SAM techniques would bridge the gap. With our present measurements we indeed fill that gap. We can also highlight that we provide noticeably lower mass limit than VLTI/PIONIER did over 50 to 100 mas.
	
Detection limits at $5\sigma$ on Fomalhaut give at least a contrast varying between $1.9\cdot 10^{-3}$ and $1.2\cdot 10^{-3}$ ($\Delta mag$ of 6 and 6.9 respectively) beyond 50\,mas.

\subsubsection*{\object{g Lupi}} 
G Lupi (also know as HD139664) is a F5V main sequence star located at 17.5\,pc \citep{vanleeuwen2007}. It is a member of the Hercules-Lyra association \citep{lopez-santiago2006}.
Dust scattered light was first observed using a coronagraph on the HST \citep{kalas2006}. It revealed a disk structure seen nearly edge-on extending out to 109 AU.

The age of g Lupi is not well constrained but various age indicators suggest that the system is 300 million years old. The uncertainties concerning the age of ${300}^{+700}_{-200}\, Myr$ imply a respective variation of $\pm 25\, M_{Jup}$ in our upper mass limits. 

Detection limits at $5\sigma$ on g Lupi give at least a contrast of $6\cdot 10^{-3}$ beyond 50\,mas, up to $2.5\cdot 10^{-3}$ at 175\,mas ($\Delta mag$ of 5.5 and 6.5 respectively).

\subsubsection*{\object{HD181327}}
HD181327 is an F5V/F6V star located at 39.4\,pc and assumed to be a member of the $\beta$ Pictoris moving group \citep{zuckerman2004}. 
It was identified as a debris disk candidate based on the detection of far-IR luminosity excess  \citep{mannings1998}. PSF-subtracted coronagraphic observations on HST data revealed a debris ring at 86 AU radius from the star \cite{schneider2006}, Herschel/PACS observations showed the presence of a Kuiper's belt-like structures populated of icy planetesimals \cite{chen2008}.

Detection limits at $5\sigma$ on HD181327 give at least a contrast of $5\cdot 10^{-3}$ beyond 50\,mas, up to $1.8\cdot 10^{-3}$ at 125\,mas ($\Delta mag$ of 5.8 and 6.8 respectively).

\subsubsection*{\object{HR8799}}
HR8799 is a member of the $\sim$30 Myr old Columba association \citep{marois2010,baines2012}. The star is an F0V star located at 39.4\,pc from Earth \citep{vanleeuwen2007}.

Our detection limits are in good agreement with those derived by \cite{hinkley2011} using the same technique at Keck telescope. This highlight the robustness of non-redundant masking techniques.
In the same way we  have done on $\beta$ Pictoris, we used \cite{marois2010} to estimate the HR8799 companion \emph{e} position. At the time of observation, we expect it to be at the position $\Delta\alpha=-0.353''$ and $\Delta\delta=-0.108''$. At these coordinates we find a corresponding upper mass boundary $m_{LOD} = 19\;M_{Jup}$ ($\Delta mag=7.4$).

Detection limits at $5\sigma$ on HR8799 give at least a contrast of $7\cdot 10^{-3}$ beyond 50\,mas, up to $2.2\cdot 10^{-3}$ ($\Delta mag$ of 5.4 and 6.7 respectively).

\begin{table}
	\caption[]{Highest contrast capabilities of SAM mode on NaCo}
	\label{SAMContrast}
	\centering
	\begin{tabular}{lcccc}
		\hline \hline \noalign{\smallskip}
		Object & $\Delta mag$ & contrast ratio & Mass & Distance \\
		&  (mag) & ($5\sigma$) & ($M_{jup}$) & (AU) \\
		\noalign{\smallskip} \hline \noalign{\smallskip}
		49 Cet & 7.86 & $7.2\times 10^{-4}$ & 17.6 & 8--12 \\
		AU Mic & 8.72 & $3.2\times 10^{-4}$ & 3.4 & 1--2\\
		$\beta$ Pic & 7.79 & $7.6\times 10^{-4}$ & 10.7 & 1--3\\
		$\eta$ Tel & 8.32  & $4.7\times 10^{-4}$ & 10.3 & 4--7\\
		Fomalhaut & 8.39 & $4.4\times 10^{-4}$ & 61.1 & 1--2\\
		g Lup & 7.44 & $1.1\times 10^{-3}$ & 24.5 & 1--3 \\
		HD181327 & 7.71 & $8.2\times 10^{-4}$ & 9.9 & 5--10\\
		HR8799 & 7.92 & $6.8\times 10^{-4}$ & 10.7 & 4--8\\
		\noalign{\smallskip} \hline \noalign{\smallskip}
	\end{tabular}
\end{table}

\begin{table}
	\caption[]{Upper mass limits: lowest mass we can safely exclude within 30 to 500 mas separation}
	\label{SAMUpperMasLimits}
	\centering
	\begin{tabular}{lcccc}
		\hline \hline \noalign{\smallskip}
		Object & $\Delta mag$ & contrast ratio & Mass \\
		&  (mag) & ($5\sigma$) & ($M_{jup}$)\\
		\noalign{\smallskip} \hline \noalign{\smallskip}
		49 Cet & 5.24 & $8.0\times 10^{-3}$ & 156.9 \\
		AU Mic & 5.67 & $5.4\times 10^{-3}$ & 10.1 \\
		$\beta$ Pic & 4.96 & $1.0\times 10^{-2}$ & 87.9 \\
		$\eta$ Tel & 5.05 & $9.6\times 10^{-3}$ & 114.1 \\
		Fomalhaut & 5.40 & $6.9\times 10^{-3}$ & 273.3 \\
		g Lup & 4.92 & $1.1\times 10^{-2}$ & 154.2 \\
		HD181327 & 5.01 & $ 9.9\times 10^{-3}$ & 73.2 \\
		HR8799 & 5.04 & $9.6\times 10^{-3}$ & 87.3 \\
		\noalign{\smallskip} \hline \noalign{\smallskip}
	\end{tabular}
\end{table}

\section{Conclusions}
\label{sectionConclusion}
The SAM technique allows for probing the very nearby environment of the star that is usually inaccessible to conventional direct imaging image processing. It combines high angular resolution and high contrast detection which are key elements in the study of planetary system formation.

While no companion has been detected in our SAM observational data, we have been able to set upper mass limits on potential companions presence within the 30 to 500\,mas range. This was done for each one of the eight debris disks reported in this study.

Especially, this data provides clarification on the presence of close companions that may have ejected known ones on more distant orbits from the star. $\beta$ Pic has an inclined orbit relatively to the main disk, this can hardly be explained otherwise than by a past dynamical interaction. HR8799 planets cannot have formed \textit{in-situ} by a core accretion process as described by \cite{pollack1996} and may have been ejected by bodies closer to the star. $\eta$ Tel B is a brown dwarf with a large separation which appears to be situated beyond the outer edge of the debris disk of its parent star, but that might have formed in the initial disk by gravitational instability \citep{cameron1978}. Fomalhaut b possesses a very eccentric orbit, that might also be explained by ejection from a previous orbit. For each object, these observations allow to exclude the presence of a range of massive bodies that might have modified the initial orbit of known companions.

The SAM technique associated with Star Hopping is particularly efficient when the correction of the wavefront is good, which is the case with NACO in $L'$. The SAM mode to be commissioned on SPHERE at the VLT will make possible to do even better in Y-J-H and K bands in a near future. This technique is recognized to be highly competitive in the brown dwarf desert exploration, and plays an important part in the study of the role of multiplicity in low masses stellar formation and the migration mechanisms.\\


\begin{acknowledgements}
	  L. Gauchet would like to thank the referee Markus Feldt for his thorough review of the paper and his constructive comments, which significantly improved the quality of the publication. 
      This work was supported by the French National Agency for Research (ANR-13-JS05-0005-01) and the European Research Council (ERC-STG-639248).
      A.-M. Lagrange acknowledges the support from the ANR blanche GIPSE (ANR-14-CE33-0018) and the Labex OSUG.
\end{acknowledgements}

\bibliographystyle{bibtex/aa.bst}
\bibliography{/home/lucien/Documents/PhD/Bibliographie/bibliography}

\begin{thebibliography}{54}
\expandafter\ifx\csname natexlab\endcsname\relax\def\natexlab#1{#1}\fi

\bibitem[{Allard {et~al.}(2012)Allard, Homeier, \& Freytag}]{allard2012}
Allard, F., Homeier, D., \& Freytag, B. 2012, Philosophical Transactions of the
  Royal Society of London A: Mathematical, Physical and Engineering Sciences,
  370, 2765, {PMID:} 22547243

\bibitem[{Aumann(1985)}]{aumann1985}
Aumann, H.~H. 1985, Publications of the Astronomical Society of the Pacific,
  97, 885

\bibitem[{Aumann {et~al.}(1984)Aumann, Beichman, Gillett, de~Jong, Houck, Low,
  Neugebauer, Walker, \& Wesselius}]{aumann1984}
Aumann, H.~H., Beichman, C.~A., Gillett, F.~C., {et~al.} 1984, The
  Astrophysical Journal, 278, L23

\bibitem[{Baines {et~al.}(2012)Baines, White, Huber, Jones, Boyajian,
  {McAlister}, Brummelaar, Turner, Sturmann, Sturmann, Goldfinger, Farrington,
  Riedel, Ireland, Braun, \& Ridgway}]{baines2012}
Baines, E.~K., White, R.~J., Huber, D., {et~al.} 2012, ApJ, 761, 57

\bibitem[{Baldwin {et~al.}(1986)Baldwin, Haniff, Mackay, \&
  Warner}]{baldwin1986}
Baldwin, J.~E., Haniff, C.~A., Mackay, C.~D., \& Warner, P.~J. 1986, Nature,
  320, 595

\bibitem[{Baraffe {et~al.}(1997)Baraffe, Chabrier, Allard, \&
  Hauschildt}]{baraffe1997}
Baraffe, I., Chabrier, G., Allard, F., \& Hauschildt, P. 1997, \aap, {arXiv:}
  astro-ph/9704144

\bibitem[{Baraffe {et~al.}(1998)Baraffe, Chabrier, Allard, \&
  Hauschildt}]{baraffe1998}
Baraffe, I., Chabrier, G., Allard, F., \& Hauschildt, P.~H. 1998, \aap, 337,
  403

\bibitem[{Baraffe {et~al.}(2003)Baraffe, Chabrier, Barman, Allard, \&
  Hauschildt}]{baraffe2003}
Baraffe, I., Chabrier, G., Barman, T.~S., Allard, F., \& Hauschildt, P.~H.
  2003, \aap, 402, 701

\bibitem[{Binks \& Jeffries(2014)}]{binks2014}
Binks, A.~S. \& Jeffries, R.~D. 2014, Monthly Notices of the Royal Astronomical
  Society: Letters, 438, L11

\bibitem[{Boccaletti {et~al.}(2015)Boccaletti, Thalmann, Lagrange, Janson,
  Augereau, Schneider, Milli, Grady, Debes, Langlois, Mouillet, Henning,
  Dominik, Maire, Beuzit, Carson, Dohlen, Engler, Feldt, Fusco, Ginski, Girard,
  Hines, Kasper, Mawet, Ménard, Meyer, Moutou, Olofsson, Rodigas, Sauvage,
  Schlieder, Schmid, Turatto, Udry, Vakili, Vigan, Wahhaj, \&
  Wisniewski}]{boccaletti2015}
Boccaletti, A., Thalmann, C., Lagrange, A., {et~al.} 2015, Nature, 526, 230

\bibitem[{Bonnefoy {et~al.}(2013)Bonnefoy, Boccaletti, Lagrange, Allard,
  Mordasini, Beust, Chauvin, Girard, Homeier, Apai, Lacour, \&
  Rouan}]{bonnefoy2013}
Bonnefoy, M., Boccaletti, A., Lagrange, A., {et~al.} 2013, Astronomy \&
  Astrophysics, 555, A107

\bibitem[{Bonnefoy {et~al.}(2014)Bonnefoy, Currie, Marleau, Schlieder,
  Wisniewski, Carson, Covey, Henning, Biller, Hinz, Klahr, Marsh~Boyer,
  Zimmerman, Janson, {McElwain}, Mordasini, Skemer, Bailey, Defrère, Thalmann,
  Skrutskie, Allard, Homeier, Tamura, Feldt, Cumming, Grady, Brandner, Helling,
  Witte, Hauschildt, Kandori, Kuzuhara, Fukagawa, Kwon, Kudo, Hashimoto,
  Kusakabe, Abe, Brandt, Egner, Guyon, Hayano, Hayashi, Hayashi, Hodapp, Ishii,
  Iye, Knapp, Matsuo, Mede, Miyama, Morino, {Moro-Martin}, Nishimura, Pyo,
  Serabyn, Suenaga, Suto, Suzuki, Takahashi, Takami, Takato, Terada, Tomono,
  Turner, Watanabe, Yamada, Takami, \& Usuda}]{bonnefoy2014}
Bonnefoy, M., Currie, T., Marleau, G., {et~al.} 2014, Astronomy \&
  Astrophysics, 562, A111

\bibitem[{Cameron(1978)}]{cameron1978}
Cameron, A. G.~W. 1978, The Moon and the Planets, 18, 5

\bibitem[{Chauvin {et~al.}(2012)Chauvin, Lagrange, Beust, Bonnefoy, Boccaletti,
  Apai, Allard, Ehrenreich, Girard, Mouillet, \& Rouan}]{chauvin2012}
Chauvin, G., Lagrange, A., Beust, H., {et~al.} 2012, Astronomy and
  Astrophysics, 542, A41

\bibitem[{Chen {et~al.}(2008)Chen, Fitzgerald, \& Smith}]{chen2008}
Chen, C.~H., Fitzgerald, M.~P., \& Smith, P.~S. 2008, The Astrophysical
  Journal, 689, 539, {arXiv:} 0808.2273

\bibitem[{Cutri {et~al.}(2003)Cutri, Skrutskie, van Dyk, Beichman, Carpenter,
  Chester, Cambresy, Evans, Fowler, Gizis, Howard, Huchra, Jarrett, Kopan,
  Kirkpatrick, Light, Marsh, {McCallon}, Schneider, Stiening, Sykes, Weinberg,
  Wheaton, Wheelock, \& Zacarias}]{cutri2003}
Cutri, R.~M., Skrutskie, M.~F., van Dyk, S., {et~al.} 2003, VizieR Online Data
  Catalog, 2246, 0

\bibitem[{Ducati(2002)}]{ducati2002}
Ducati, J.~R. 2002, VizieR Online Data Catalog, 2237, 0

\bibitem[{Gray \& Corbally(2014)}]{gray2014}
Gray, R.~O. \& Corbally, C.~J. 2014, The Astronomical Journal, 147, 80

\bibitem[{Gray {et~al.}(2006)Gray, Corbally, Garrison, {McFadden}, Bubar,
  {McGahee}, {O’Donoghue}, \& Knox}]{gray2006}
Gray, R.~O., Corbally, C.~J., Garrison, R.~F., {et~al.} 2006, \aj, 132, 161

\bibitem[{Greenbaum {et~al.}(2015)Greenbaum, Pueyo, Sivaramakrishnan, \&
  Lacour}]{greenbaum2015}
Greenbaum, A.~Z., Pueyo, L., Sivaramakrishnan, A., \& Lacour, S. 2015, ApJ,
  798, 68

\bibitem[{Haniff {et~al.}(1987)Haniff, Mackay, Titterington, Sivia, Baldwin, \&
  Warner}]{haniff1987}
Haniff, C.~A., Mackay, C.~D., Titterington, D.~J., {et~al.} 1987, Nature, 328,
  694

\bibitem[{Hinkley {et~al.}(2011)Hinkley, Carpenter, Ireland, \&
  Kraus}]{hinkley2011}
Hinkley, S., Carpenter, J.~M., Ireland, M.~J., \& Kraus, A.~L. 2011, The
  Astrophysical Journal Letters, 730, L21

\bibitem[{Holmberg {et~al.}(2009)Holmberg, Nordström, \&
  Andersen}]{holmberg2009}
Holmberg, J., Nordström, B., \& Andersen, J. 2009, Astronomy and Astrophysics,
  501, 941

\bibitem[{Kalas {et~al.}(2008)Kalas, Graham, Chiang, Fitzgerald, Clampin, Kite,
  Stapelfeldt, Marois, \& Krist}]{kalas2008}
Kalas, P., Graham, J.~R., Chiang, E., {et~al.} 2008, Science, 322, 1345,
  {arXiv:} 0811.1994

\bibitem[{Kalas {et~al.}(2005)Kalas, Graham, \& Clampin}]{kalas2005}
Kalas, P., Graham, J.~R., \& Clampin, M. 2005, Nature, 435, 1067

\bibitem[{Kalas {et~al.}(2006)Kalas, Graham, Clampin, \&
  Fitzgerald}]{kalas2006}
Kalas, P., Graham, J.~R., Clampin, M.~C., \& Fitzgerald, M.~P. 2006, ApJ, 637,
  L57

\bibitem[{Kalas {et~al.}(2004)Kalas, Liu, \& Matthews}]{kalas2004}
Kalas, P., Liu, M.~C., \& Matthews, B.~C. 2004, Science, 303, 1990, {PMID:}
  14988511

\bibitem[{Kennedy \& Kenyon(2008)}]{kennedy2008}
Kennedy, G.~M. \& Kenyon, S.~J. 2008, ApJ, 682, 1264

\bibitem[{Kenworthy {et~al.}(2013)Kenworthy, Meshkat, Quanz, Girard, Meyer, \&
  Kasper}]{kenworthy2013}
Kenworthy, M.~A., Meshkat, T., Quanz, S.~P., {et~al.} 2013, ApJ, 764, 7

\bibitem[{Lacour {et~al.}(2011)Lacour, Tuthill, Amico, Ireland, Ehrenreich,
  Huelamo, \& Lagrange}]{lacour2011}
Lacour, S., Tuthill, P., Amico, P., {et~al.} 2011, \aap, 532, A72

\bibitem[{Lagrange {et~al.}(2010)Lagrange, Bonnefoy, Chauvin, Apai, Ehrenreich,
  Boccaletti, Gratadour, Rouan, Mouillet, Lacour, \& Kasper}]{lagrange2010}
Lagrange, A., Bonnefoy, M., Chauvin, G., {et~al.} 2010, Science, 329, 57,
  {PMID:} 20538914

\bibitem[{Lohmann {et~al.}(1983)Lohmann, Weigelt, \& Wirnitzer}]{lohmann1983}
Lohmann, A.~W., Weigelt, G., \& Wirnitzer, B. 1983, Appl. Opt., 22, 4028

\bibitem[{{López-Santiago} {et~al.}(2006){López-Santiago}, Montes,
  {Crespo-Chacón}, \& {Fernández-Figueroa}}]{lopez-santiago2006}
{López-Santiago}, J., Montes, D., {Crespo-Chacón}, I., \&
  {Fernández-Figueroa}, M.~J. 2006, ApJ, 643, 1160

\bibitem[{Mamajek(2012)}]{mamajek2012}
Mamajek, E.~E. 2012, \apj, 754, L20

\bibitem[{Mannings \& Barlow(1998)}]{mannings1998}
Mannings, V. \& Barlow, M.~J. 1998, ApJ, 497, 330

\bibitem[{Marois {et~al.}(2010)Marois, Zuckerman, Konopacky, Macintosh, \&
  Barman}]{marois2010}
Marois, C., Zuckerman, B., Konopacky, Q.~M., Macintosh, B., \& Barman, T. 2010,
  Nature, 468, 1080

\bibitem[{Nielsen {et~al.}(2014)Nielsen, Liu, Wahhaj, Biller, Hayward, Males,
  Close, Morzinski, Skemer, Kuchner, Rodigas, Hinz, Chun, Ftaclas, \&
  Toomey}]{nielsen2014}
Nielsen, E.~L., Liu, M.~C., Wahhaj, Z., {et~al.} 2014, ApJ, 794, 158

\bibitem[{Perryman {et~al.}(1997)Perryman, Lindegren, Kovalevsky, Hoeg,
  Bastian, Bernacca, Crézé, Donati, Grenon, Grewing, van Leeuwen, van~der
  Marel, Mignard, Murray, Le~Poole, Schrijver, Turon, Arenou, Froeschlé, \&
  Petersen}]{perryman1997}
Perryman, M. A.~C., Lindegren, L., Kovalevsky, J., {et~al.} 1997, Astronomy and
  Astrophysics, 323, L49

\bibitem[{Plavchan {et~al.}(2009)Plavchan, Werner, Chen, Stapelfeldt, Su,
  Stauffer, \& Song}]{plavchan2009}
Plavchan, P., Werner, M.~W., Chen, C.~H., {et~al.} 2009, \apj, 698, 1068

\bibitem[{Pollack {et~al.}(1996)Pollack, Hubickyj, Bodenheimer, Lissauer,
  Podolak, \& Greenzweig}]{pollack1996}
Pollack, J.~B., Hubickyj, O., Bodenheimer, P., {et~al.} 1996, Icarus, 124, 62

\bibitem[{Roberge {et~al.}(2013)Roberge, Kamp, Montesinos, Dent, Meeus,
  Donaldson, Olofsson, Moór, Augereau, Howard, Eiroa, Thi, Ardila, Sandell, \&
  Woitke}]{roberge2013}
Roberge, A., Kamp, I., Montesinos, B., {et~al.} 2013, ApJ, 771, 69

\bibitem[{Schneider {et~al.}(2006)Schneider, Silverstone, Hines, Augereau,
  Pinte, Ménard, Krist, Clampin, Grady, Golimowski, Ardila, Henning, Wolf, \&
  Rodmann}]{schneider2006}
Schneider, G., Silverstone, M.~D., Hines, D.~C., {et~al.} 2006, \apj, 650, 414

\bibitem[{Smith \& Terrile(1984)}]{smith1984}
Smith, B.~A. \& Terrile, R.~J. 1984, Science, 226, 1421, {PMID:} 17788996

\bibitem[{Spiegel \& Burrows(2012)}]{spiegel2012}
Spiegel, D.~S. \& Burrows, A. 2012, ApJ, 745, 174

\bibitem[{Stauffer {et~al.}(2010)Stauffer, Tanner, Bryden, Ramirez, Berriman,
  Ciardi, Kane, Mizusawa, Payne, Plavchan, Braun, Wyatt, \&
  Kirkpatrick}]{stauffer2010}
Stauffer, J., Tanner, A.~M., Bryden, G., {et~al.} 2010, \pasp, 122, 885

\bibitem[{Tokunaga(2000)}]{tokunaga2000}
Tokunaga, A.~T. 2000, in Allen's astrophysical quantities, 4th edition, 143,
  ed. A. N. Cox

\bibitem[{Torres {et~al.}(2006)Torres, Quast, da~Silva, de~la Reza, Melo, \&
  Sterzik}]{torres2006}
Torres, C. A.~O., Quast, G.~R., da~Silva, L., {et~al.} 2006, Astronomy and
  Astrophysics, 460, 695

\bibitem[{Tuthill {et~al.}(2000)Tuthill, Monnier, Danchi, Wishnow, \&
  Haniff}]{tuthill2000}
Tuthill, P., Monnier, J., Danchi, W., Wishnow, E., \& Haniff, C. 2000,
  Publications of the Astronomical Society of the Pacific, 112, 555

\bibitem[{van Leeuwen(2007)}]{vanleeuwen2007}
van Leeuwen, F. 2007, Astronomy and Astrophysics, 474, 653

\bibitem[{Weigelt(1977)}]{weigelt1977}
Weigelt, G.~P. 1977, Optics Communications, 21, 55

\bibitem[{Wyatt {et~al.}(2007)Wyatt, Smith, Su, Rieke, Greaves, Beichman, \&
  Bryden}]{wyatt2007}
Wyatt, M.~C., Smith, R., Su, K. Y.~L., {et~al.} 2007, \apj, 663, 365

\bibitem[{Zuckerman \& Song(2004)}]{zuckerman2004}
Zuckerman, B. \& Song, I. 2004, \apj, 603, 738

\bibitem[{Zuckerman \& Song(2012)}]{zuckerman2012}
Zuckerman, B. \& Song, I. 2012, \apj, 758, 77

\bibitem[{Zuckerman {et~al.}(2001)Zuckerman, Song, Bessell, \&
  Webb}]{zuckerman2001}
Zuckerman, B., Song, I., Bessell, M.~S., \& Webb, R.~A. 2001, \apj, 562, L87

\end{thebibliography}

\end{document}